# Reconfigurable cascaded thermal neuristors for neuromorphic computing

Erbin Qiu[1,2,*], Yuan-Hang Zhang[2], Massimiliano Di Ventra[2] and Ivan K. Schuller[2]

**Affiliations:**

[1] Department of Electrical and Computer Engineering, University of California San Diego; La Jolla, CA 92093, USA.

[2] Department of Physics, University of California San Diego; La Jolla, CA 92093, USA.

*Corresponding author: Erbin Qiu e3qiu@eng.ucsd.edu

**Abstract:**

While the complementary metal-oxide semiconductor (CMOS) technology is the mainstream for the hardware implementation of neural networks, we explore an alternative route based on a new class of spiking oscillators we call "thermal neuristors", which operate and interact solely via thermal processes. Utilizing the insulator-to-metal transition in vanadium dioxide, we demonstrate a wide variety of reconfigurable electrical dynamics mirroring biological neurons. Notably, inhibitory functionality is achieved just in a single oxide device, and cascaded information flow is realized exclusively through thermal interactions. To elucidate the underlying mechanisms of the neuristors, a detailed theoretical model is developed, which accurately reflects the experimental results. This study establishes the foundation for scalable and energy-efficient thermal neural networks, fostering progress in brain-inspired computing.

**One-Sentence Summary:**

We achieved cascaded information flow and inhibitory functionality in spiking neurons solely through thermal interactions.

# Main text:

Neuromorphic computing, which takes inspiration from the brain's information processing capabilities, offers an energy-efficient alternative to traditional von Neumann architectures (*1–6*). At the heart of neuromorphic computing are spiking neural networks (SNNs) (*1, 7*), which simulate the event-driven nature and sparse communication patterns of biological neurons by using precisely timed spikes across layers of artificial neurons and synapses. Input data is represented and transmitted through time-varying spikes that are processed by interconnected neurons. Much of the current research has been centered on software simulations (*8, 9*) or implementations using complementary metal-oxide-semiconductors (CMOS) (*10–13*). Notable CMOS-based SNNs include IBM's TrueNorth chip (*11*) and Intel's Loihi (*13*) which are built using cutting-edge, costly technologies and complex circuit designs. A CMOS neuron typically encompasses components like temporal integration, spike/event generation, refractory period, spike frequency adaptation, and spiking threshold adaptation blocks (*12*). However, the significant circuit footprint, limitations in scaling, and energy consumption may impede the progress of CMOS-based SNNs.

Beyond CMOS-based models, there has been a recent emergence of spiking neuron devices (*14, 15*) constructed using quantum materials, which are now at the forefront of neuromorphic computing, including but not limited to Mott neurons (*16–19*), magnetic neurons (*20–22*) and phase change neurons (*23–25*). These devices have the potential to significantly reduce both the circuit complexity and the physical size of artificial neurons. However, the development of these quantum material-based artificial spiking neuron devices is still in its infancy, and various challenges and issues have been observed in initial demonstrations.

One of the primary challenges facing artificial spiking neuron devices, which is often overlooked, is the difficulty in directly transmitting information between layers without intricate circuit configurations. The issue stems from the fact that the presence of a subsequent neural layer alters the output of the preceding layer due to the loading effect. Solutions typically involve the integration of complex buffer circuits (*26, 27*), which substantially increase the overall size, often overshadowing the spiking neurons themselves in terms of space. Some studies sidestep this issue altogether (*28–30*), focusing solely on network-level simulations based on the properties of individual neuron devices, without considering the challenges of transmitting information between layers at hardware level. As a result, the efficient integration of cascading neural layers remains elusive.

Furthermore, current spiking neuron configurations lack versatility. For instance, inhibitory neurons play a critical role in neural activities, but replicating this functionality in artificial spiking neurons is no easy feat. Some attempted solutions involve elaborate circuits (*31–33*), synaptic weight alternations (*26*), or optical inhibition (*27*). However, none of these can directly implement an inhibitory neuron in a single simple device, which poses significant constraints on the application of learning algorithms.

In our study, we introduce a dynamic system comprised of two thermally coupled spiking oscillators based on Mott insulators, which effectively addresses the aforementioned challenges. These spiking oscillators, referred to as neuristors, exhibit a range of neural functions. Notably, we demonstrate the implementation of an inhibitory neuristor using simple Mott oxides, such as $VO_2$, by trapping the metallic state, eliminating the need for complex circuits. Additionally, both excitatory and inhibitory neuristors can be realized using the same device by employing different inputs, thereby enhancing the device's versatility and applicability. The neuristor also displays a rich array of reconfigurable electrical behaviors such as rate coding and stochastic leaky integrate-and-fire. Crucially, we demonstrate the feasibility of cascading neural layers through thermal interactions, which effectively eliminate the necessity for complex input/output circuits between layers. Our straightforward and innovative approach paves the way for advancements in reconfigurable cascading neural layers, which hold promise for applications in artificial intelligence.

## Results

We have successfully implemented neuristors with reconfigurable functionalities using thermally coupled spiking oscillators, which are based on the insulator-to-metal transition (IMT) of the Mott insulator $VO_2$. We patterned a 100 nm thick $VO_2$ thin film into an array of nanodevices, each measuring 100 x 500 nm². As illustrated in Fig. 1A, each nanodevice is separated from its neighbors by a 500 nm gap. To examine the thermal interactions between neuristors, we etched away the $VO_2$ material between the nanodevices to electrically isolate each spiking oscillator. Despite electrical isolation, the oscillators remain thermally coupled through the $Al_2O_3$ substrate, which serves as an effective thermal conductor due to its high thermal conductivity.

The working principle of the spiking oscillator circuit is described as follows. As depicted in Fig. 1B, the $VO_2$ nanodevice, initially in its insulating state, is connected in series with a variable load resistor, $R_{load}$, and in parallel with an intrinsic parasitic capacitance, C. When an input voltage is applied to the circuit, the ensuing current heats up the $VO_2$ nanodevice, concurrently charging up the parasitic capacitance. Upon reaching the critical threshold voltage, the $VO_2$ undergoes an IMT. This abrupt decrease in the $VO_2$ resistance prompts the parasitic capacitance to discharge, resulting in a current spike. As the capacitance discharges, most input voltage is dropped across $R_{load}$, and the voltage across $VO_2$ does not generate sufficient heat to sustain the metallic state, causing the $VO_2$ to revert to its insulating state (*16, 17, 34*). This process repeats, generating a series of stable, spiking, auto-oscillations as depicted in Fig. S1. By tuning the input voltage and load resistance, a wide variety of reconfigurable neural dynamics can be demonstrated within the same neuristor.

### *Single neuristor characteristics*

First, we demonstrate that our Mott insulator $VO_2$-based neuristor exhibits spiking behavior analogous to a biological neuron. Specifically, we demonstrate the all-or-nothing law (*34*), which states that with external stimuli, a neuristor either gives maximal response or no response at all; and the rate coding law (*17, 35*), showing that the spiking frequency increases with increasing input stimulus.

In Fig. 2, we conducted measurements on a single neuristor configured in series with a 12 kΩ load resistor at a base temperature of 325 K under varying input voltages. With a 9 V bias, the neuristor displays no response due to the subthreshold input, as illustrated in Fig. 2A. With a suprathreshold voltage of 12.5 V, the neuristor exhibits a full current spiking response as depicted in Fig. 2B. Almost all current spikes exhibit nearly identical amplitudes, except the first, which has a larger amplitude due to the initial insulating state requiring a higher firing threshold, causing a larger current surge. This behavior, also observed at different suprathreshold voltages (Fig. S2), exemplifies the "all-or-nothing" law (*34*).

Although the amplitude of the current spikes remains constant as the input voltage varies, there is an observed increase in spiking frequency with increasing input voltage, as shown in Fig. S2. We plotted the spiking frequencies against each input voltage (Fig. 2D), revealing a discontinuous frequency-voltage relationship with a threshold at 10.5 V. Beyond this, the firing rate increases with stimulus intensity, mirroring typical type-II neuronal rate coding (*27, 36, 37*).

However, our neuristor exhibits a critical feature differing from traditional neuron models, which is the key to implementing the inhibitory functionality. As demonstrated in Fig. 2C, when biased with 15.8 V, the neuristor becomes quiescent after a few initial spikes. This phenomenon arises because $VO_2$ gets trapped in its metallic state, with a constant current flowing through the metallic filament, instead of reverting to its insulating state (*38, 39*). The inhibitory behavior is also reproduced with a bias voltage of 17 V, as depicted in Fig. S2K. To rule out the possibility of nanodevice degradation causing this behavior, we tested at 15.7

V again and confirmed stable spiking behavior, as shown in Fig. S2L. We incorporated this distinctive inhibitory feature into the rate coding graph in Fig. 2D, presenting a comprehensive representation of neuristor behavior.

To understand the mechanism of the spiking oscillators, we constructed a theoretical model, which comprises of the circuit from Fig. 1B, the hysteresis model in Fig. 1C, and a simple heat conduction model. As depicted in Fig. 2A-D, the simulations align well with the experimental results. A detailed explanation of this model is presented in the supplemental materials (SM).

Using the theory model, we show further insights into the three distinct operational modes of our neuristor, depicted in the resistance-time and resistance-temperature plots in Fig. 2E. At a subthreshold voltage of 9 V, the generated heat is insufficient to trigger the IMT, thereby confining $VO_2$ to its insulating state. Conversely, at 15.8 V, the system produces excessive heat, preventing $VO_2$ from fully reverting to its insulating state before the arrival of the next current spike, eventually trapping $VO_2$ in its metallic state. However, with a moderate voltage of 12.5V, $VO_2$ is adequately heated to transition into its metallic state and subsequently allowed sufficient time to fully revert to its insulating state. This allows $VO_2$ to traverse the full hysteresis loop and return to its initial state, facilitating stable oscillations.

## *Reconfigurable neural functionalities*

The diverse reconfigurable neural functionalities are manifested via the thermal interaction of coupled neuristors. As shown in the schematic inset of Fig. 3A, the two neuristors, spaced 500 nm apart, are electrically isolated but thermally coupled through the $Al_2O_3$ substrate. The phenomenon of thermal interactions between spiking oscillators has been previously documented (*40*, *41*). Leveraging the characteristics of a single neuristor, the pair of thermally coupled neuristors exhibits a diverse range of reconfigurable neural functionalities, which can be tuned by adjusting their input voltages and load resistances.

In Fig. 3A, neuristor A is subjected to a short 200 ns pulse at 1.3 V without any load resistor, generating a 4 mA current pulse. This, in turn, creates a heat spike that propagates to neuristor B, which locally increases neuristor B's temperature, lowers its threshold voltage and causes the IMT at a subthreshold voltage of 1.5 V. Since neuristor B is not connected to any load resistor, it stays in the metallic state and yields a direct current (DC) output.

The spike-in and DC-out effect, as shown in Fig. 3A, can be reconfigured to a spike-in and spike-out behavior, as shown in Fig. 3B, by incorporating a 9 kΩ load resistor to neuristor B. On its own, with a 2.7 V bias and no thermal interaction, neuristor B is unable to trigger spikes, as illustrated in Fig. S3. However, when neuristor A is subjected to a brief 200 ns pulse at 3.3 V, the heat spike it generates enables neuristor B to produce stable spikes. Remarkably, this process is highly energy efficient. As an example, a single spike consuming 6.45 nJ from neuristor A can initiate 14 spikes in neuristor B with a total energy output of 5.56 nJ, as detailed in Fig. S4.

The electrical dynamics of coupled neuristors can also be reconfigured to exhibit another distinctive feature, known as stochastic leaky integrate-and-fire. In this configuration, neuristor A is consistently subjected to a suprathreshold voltage of 7V in series with a 22 kΩ load resistor, resulting in stable spikes, while neuristor B, in series with a 30 kΩ load resistor, is subjected to a subthreshold voltage of 5 V, which alone is not sufficient for it to fire. However, neuristor B can integrate multiple current spikes from neuristor A and fire a spike, as shown in Fig. 3C. The initial current spike from neuristor A is insufficient to elevate the temperature enough to activate neuristor B. Yet, the accumulated heat from multiple current spikes eventually causes neuristor B to undergo the IMT, resulting in a spike (*17*). Importantly, the number of spikes neuristor B integrates from neuristor A before firing is inherently stochastic due to jittering behaviors in the neuristor (*42*) and stochasticity in thermal propagation (*43*). In this case, neuristor B may integrate 3, 4, 5, or 8 spikes from neuristor A to produce a single spike.

Fig. 3D depicts how the stochastic firing probability of the number of integrated spikes from neuristor A changes with different subthreshold voltages applied to neuristor B. When neuristor B is subjected to stronger subthreshold stimuli, it exhibits a more deterministic firing probability and requires fewer integrated spikes from neuristor A. Conversely, weaker subthreshold stimuli result in a more pronounced stochastic leaky integrate-and-fire behavior (*17*). For a more comprehensive and detailed view of the stochastic and deterministic spiking patterns, Fig. S5 presents the behavior of two thermally coupled neuristors, highlighting the distinctive stochastic leaky integrate-and-fire characteristic. Distance between neuristors also affects their coupling pattern, and a demonstration is shown in Fig. S6.

To the best of our knowledge, this is the first instance of cascading neural layers in hardware exclusively implemented with thermally coupled neuristors, eliminating the need for complicated CMOS circuits. The top section of Fig. 4 shows a flowchart that illustrates the information transfer process between the neural layers. Neuristor A integrates multiple input electrical pulses and produces a current spike, which acts as the heat spike input for neuristor B through the sapphire substrate. Subsequently, neuristor B stochastically integrates the propagated heat spike from neuristor A to create its own spike.

As depicted in the bottom left panel, neuristor A in the preceding layer is supplied with consecutive square pulses at 9.1 V, with a period of 1μs and a 50% duty cycle (light blue curve). The rapid charging and discharging induced by these square pulses result in a serrated voltage curve for neuristor A (black curve). When the accumulated voltage across neuristor A reaches the threshold, it undergoes an IMT, leading to a full discharge and the generation of a current spike. Simultaneously, while neuristor B in the subsequent layer is subjected to a subthreshold DC voltage of 5.7 V, the heat spike from neuristor A lowers the threshold voltage and triggers an IMT in neuristor B (red curve), also resulting in a spike (see Fig. S7). In this way, we achieve cascaded information transfer between different neural layers at the hardware level , eliminating the necessity for complex input/output circuits between layers (*26*).

The bottom right panel portrays a similar scenario, but with input square pulses having an 80% duty cycle. Due to the extended duty cycle, neuristor A has longer charging and shorter discharging time per pulse. This means it needs fewer input pulses to reach the threshold, and consequently fires a spike at a faster rate. However, this does not necessarily speed up spiking in neuristor B, as its charging time to reach the threshold remains invariant. As a result, some current spikes from neuristor A do not trigger a corresponding spike from neuristor B, adhering to its refractory period – another key neural function (*17*, *19*). As depicted in the bottom right panel, neuristor B needs to integrate 2 or 3 spikes from neuristor A to generate its own spike. More details about the impact of the duty cycle and the pulse amplitude can be found in Fig. S7 and Fig. S8. By adjusting the input waveform, the cascaded neuristors exhibit rich reconfigurable dynamics, effectively modulating the flow of information.

In large-scale spiking neural networks, not only is the information transfer important, but the inhibitory functionality is also significant (*7*, *9*, *26*). In Fig. 5, we demonstrate the versatility of our neuristors, where excitatory and inhibitory functionalities are displayed within the same neuristor by controlling their input voltages.

An excitatory neuristor becomes active upon receiving external stimuli. In our case, the neuristors interact with each other through heat interaction via the sapphire substrate. Neuristor A functions as a heat pump, generating a continuous stream of heat spikes when biased at 2.9 V. Neuristor B, with an applied subthreshold voltage of 2.6 V, is excited by neuristor A in a 1:1 excitation mode, as shown in Fig. 5A. The heat synchronizes the phases and frequencies of the two neuristors. When neuristor A is subjected to suprathreshold voltage of 4.3 V while neuristor B remains at a subthreshold voltage of 2.6 V, neuristor A excites neuristor B in a mixed integer excitation mode with 3:1 and 2:1 spiking pattern, demonstrating stochastic leaky-and-fire characteristics. Notably, the heat still maintains phase-locking between both neuristors. The evolution of excitatory interaction characteristics in response to changes in input voltage is shown in Fig. S9.

We can convert neuristor B from an excitatory to an inhibitory neuristor by carefully adjusting its bias voltage just below its upper limit. Surprisingly, the inhibitory functionality emerges in the same device without any physical alterations. As shown in Fig. 5D, neuristor B exhibits stable spiking when neuristor A is inactive. However, upon neuristor A's activation, neuristor B is effectively inhibited, ceasing to spike after a few initial spikes. The stream of heat spikes from neuristor A raises the temperature of neuristor B and this pushes B above the upper threshold of stable rate coding as described in Fig. 2D. This consequently traps neuristor B in the metallic state with a conductive filament and disables its capability to sustain stable auto-oscillations. Another reconfigurable inhibitory characteristic is presented in Fig. S10, showing mutual inhibition between neuristors.

By introducing a thermal coupling term for adjacent neuristors, our theoretical model accurately replicates these behaviors, as shown in Fig. 5C and 5F. The SM provides more details of this model, including a study on the effect of coupling strength in Fig. S11, and additional simulations of the excitatory and inhibitory behaviors in Figs. S12 and S13.

**Conclusion**

In this study, we engineered and analyzed thermally coupled neuristors utilizing the insulator-to-metal transition of vanadium dioxide. By exploiting $VO_2$'s hysteresis loop and thermal interactions, we demonstrated versatile neural dynamics without relying on complex CMOS circuits. This paves the way for scalable spiking neural networks and computation blocks like logic gates and feed-forward layers. Furthermore, our comprehensive theoretical model elucidates the neuristors' operational principles, facilitating simulation and design of large-scale neuristor networks. This opens avenues for the advancement of efficient and compact neural networks with applications spanning artificial intelligence to brain-inspired computing.

Nonetheless, our design has its limitations. The spatial layout imposes constraints on thermal interactions, which in turn limit the fan-in/fan-out capacities of neuristors. Moreover, regulating heat flows in large-scale networks presents an important challenge. Intriguingly, a large, unregulated neuristor network could potentially be used as a reservoir in reservoir computing. Moreover, investigating long-range correlations among distantly positioned neuristors could also yield fascinating insights into the brain's dynamic behavior.

**Acknowledgments:**

**Funding:**

E.Q. and I.K.S. were supported by the Air Force Office of Scientific Research under award number FA9550-22-1-0135. Y.-H.Z. and M.D. were supported by the Department of Energy under Grant No. DE-SC0020892.

**Author contributions:**

E.Q. and I.K.S. conceived and designed the project. E.Q. fabricated the samples and performed all the measurements. Y.-H. Z. performed the modelling simulations. All authors participated in the discussion and interpretation of the results. E.Q., Y.-H. Z., M.D. and I.K.S. wrote the manuscript. I.K.S. and M.D. supervised the project.

**Competing interests:**

Authors declare that they have no competing interests.

**Data and materials availability:** All data is available in the manuscript or the supplementary materials. All data, code, and materials can be requested from the authors.

# Figures

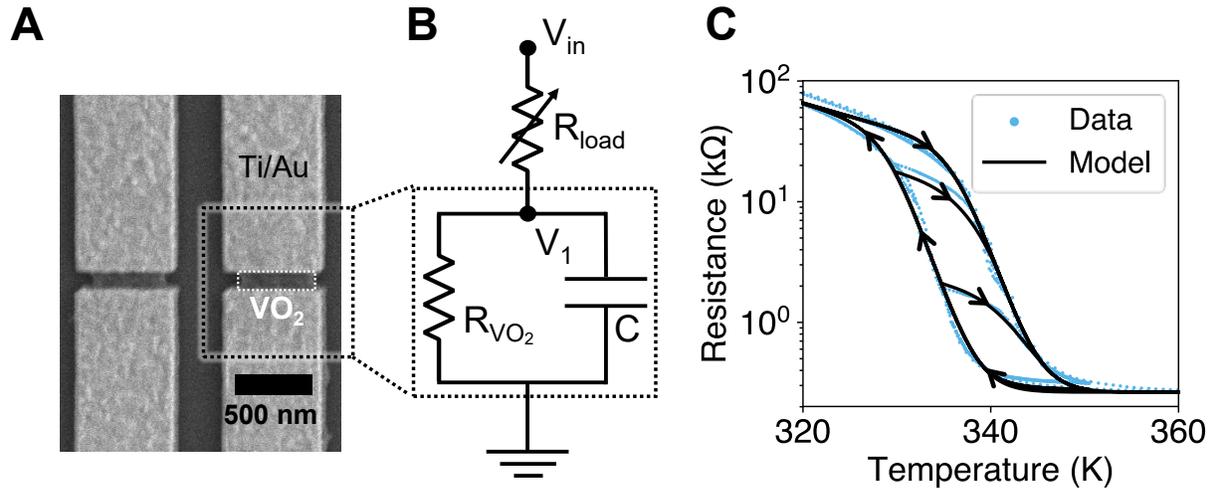

**Fig. 1. Spiking oscillators as thermally coupled neuristors.** (A) An SEM image of two adjacent $VO_2$ nanodevices. Each nanodevice has dimensions of 100 x 500 nm². To study the thermal interactions between two spiking oscillators, the nanodevices are placed in close proximity, separated by a 500 nm gap, electrically isolated by etching away the $VO_2$ between them. (B) The schematic of the equivalent circuit setup for a single spiking oscillator. To generate electrical spikes, a variable load resistor, $R_{load}$, is connected in series with the $VO_2$ nanodevice. An intrinsic parasitic capacitance, approximately 0.15 nF, is connected in parallel with the $VO_2$ nanodevice. The circuit represents a single spiking oscillator configured as a neuristor, which is reconfigured for various neuronal functionalities in subsequent measurements. (C) The hysteresis loops of resistance versus temperature for both experimental data and theoretical modeling. The major loop ranges from 320 K to 360 K and two minor loops start from 330 K and 335 K, respectively. The cooling branches of the loops overlap, while the heating branches vary for different loops. The major and minor loops are essential for realizing excitatory and inhibitory neuronal functionalities. The theoretical model closely aligns with the experimental results.

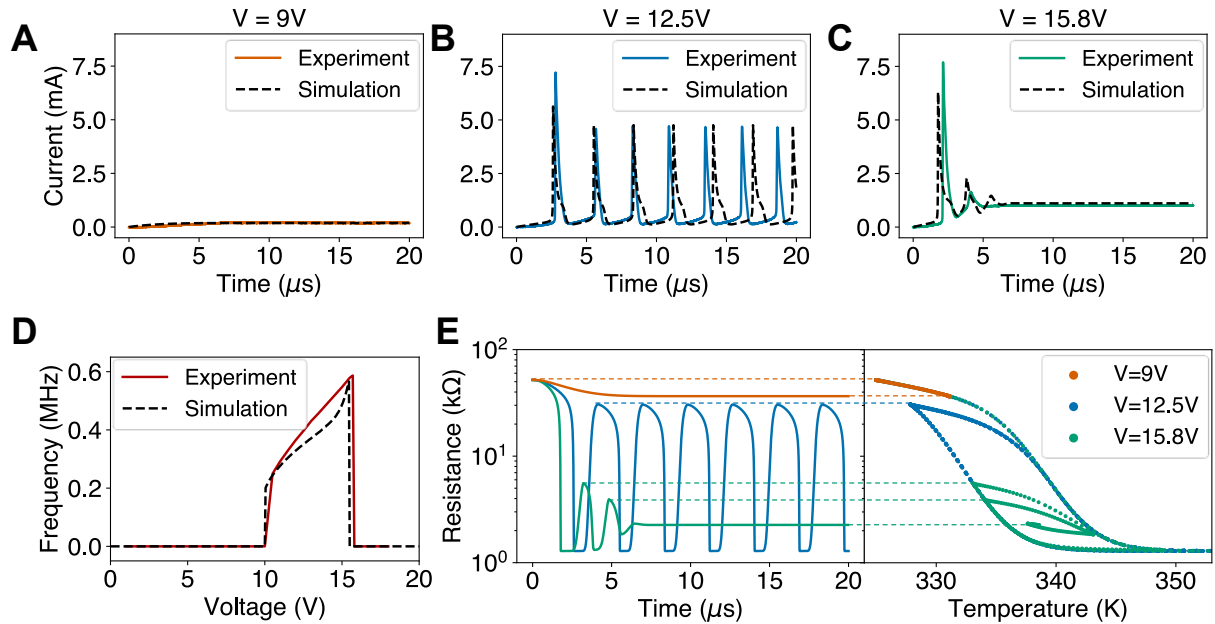

**Fig. 2. Single neuristor characteristics.** All measurements in this figure were conducted at a base temperature of 325 K, with a load resistance of 12 kΩ. (A) With an applied voltage of 9 V, the neuristor shows no current response in either the experiment (solid line) or simulation (dashed line). (B) When subjected to 12.5 V, the neuristor exhibits a full current response in both the experiment (solid line) and simulation (dashed line). This phenomenon is termed the "all-or-nothing" law. (C) When subjected to 15.8 V, both the experimental (solid line) and simulated (dashed line) neuristor become inhibited after the initial two spikes, failing to return to the insulating state and maintaining a thin metallic filament through which a small (1mA) current flow. (D) The graph shows the spiking frequency as a function of input voltages, for both experimental (solid line) and simulation (dashed line) data. This typical type-II neuronal functionality shows a discontinuous frequency jump upon reaching the 10.5 V threshold voltage. In between the threshold and cutoff voltages, the spiking frequency changes monotonically, a phenomenon known as "rate coding". (E) Simulated resistance versus time (left panel), and resistance as a function of temperature (right panel) at different input voltages. This illustrates the mechanism behind the three different working modes of our neuristor.

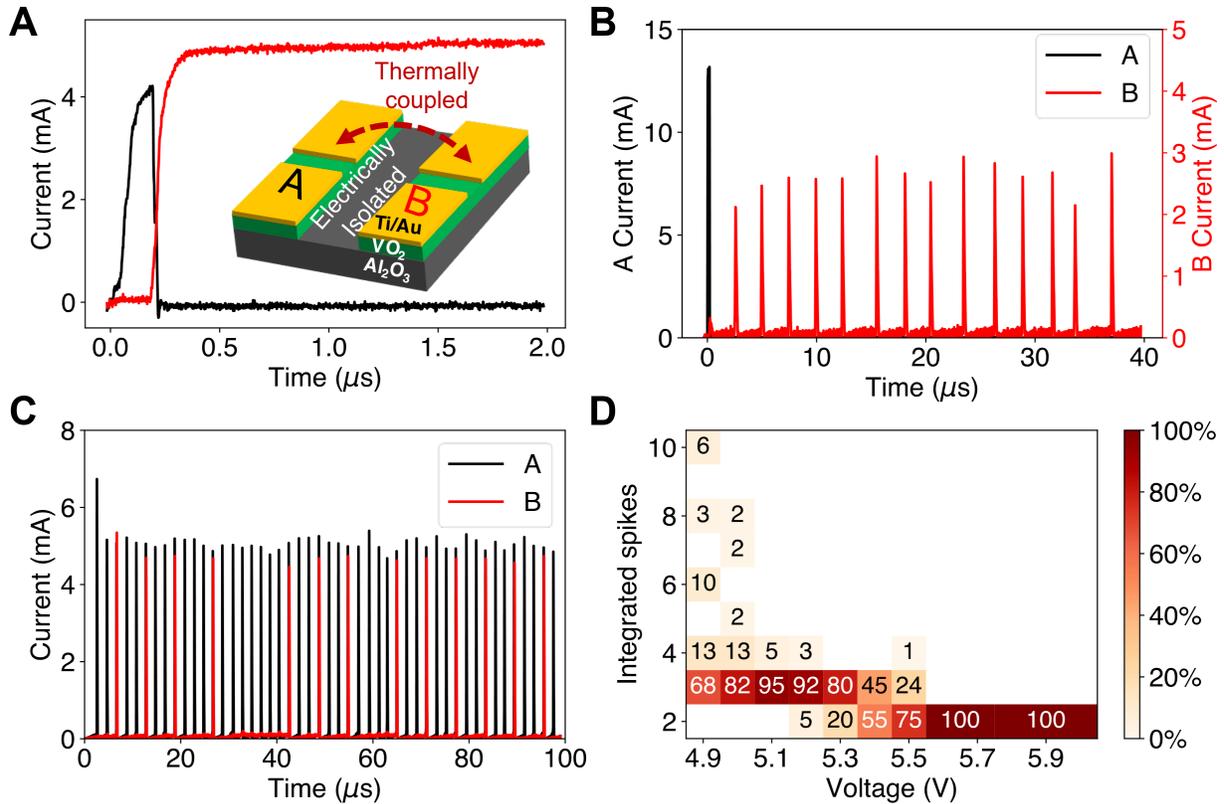

**Fig. 3. Reconfigurable electrical dynamics in coupled neuristors.** (A) Spike-in and DC-out. The load resistance for both neuristors is set to zero in this configuration. Neuristor A (black), triggered by a 1.3 V pulse lasting 200 ns, generates a 4 mA current spike. This functions as a heat spike-in for neuristor B (red), which is biased at a subthreshold voltage of 1.5 V. This induces an IMT and leads to a DC output. This is termed as the spike-in and DC-out effect. (B) Spike-in and spike-out. By adding a 9 kΩ load resistor to neuristor B, the output of neuristor B can be reconfigured from DC-out to spike-out. (C) Stochastic leaky integrate-and-fire functionality. This is an example where neuristor A, with a suprathreshold input voltage of 7V, generates stable spikes, and neuristor B, subjected to a 5V subthreshold voltage, produces current spikes by integrating multiple current spikes from neuristor A. (D) Heatmap of a comprehensive stochastic leaky integrate-and-fire behavior. This demonstrates the relationship between firing probability and input voltages. The x-axis represents the input subthreshold voltage to neuristor B, and the y-axis shows the number of spikes integrated from neuristor A. The heatmap values indicate the firing probabilities (in percentages) of neuristor B integrating different numbers of spikes for various input voltages.

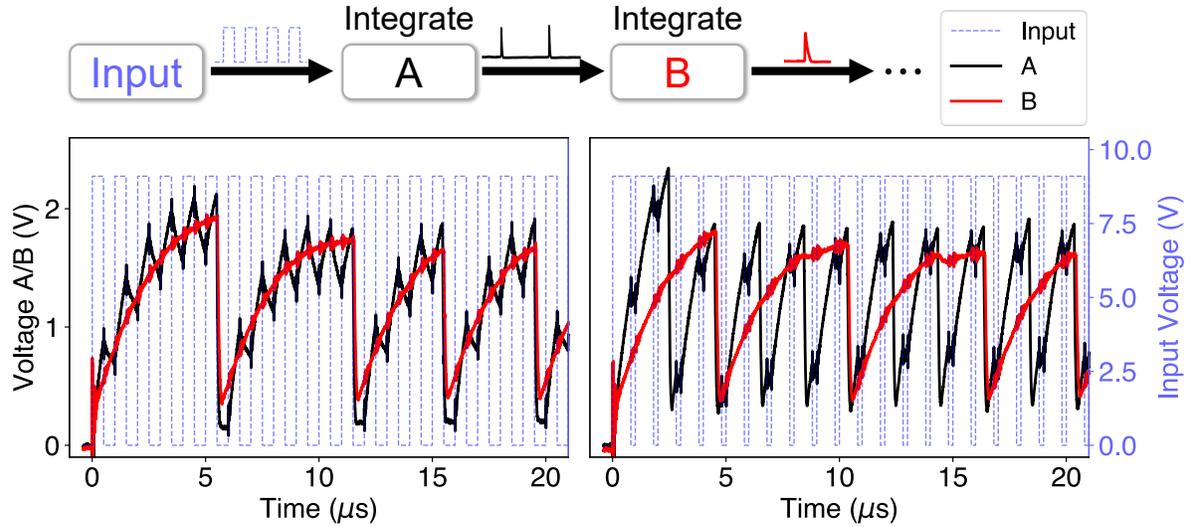

Fig. 4. **Cascaded information transfer between different neural layers**. Both neuristors A and B are connected in series with load resistors – 22 kW for neuristor A and 30 kW for neuristor B – and share the same input threshold voltage of 6 V. Neuristor A is fed with a sequence of suprathreshold pulses at 9.1 V, each with a 1μs period, while neuristor B is powered with a constant subthreshold voltage of 5.7 V. Top: Flowchart illustrating the cascaded information transfer between different neural layers via thermal interactions through $Al_2O_3$ substrate. Neuristor A in the preceding layer integrates multiple electrical pulses from the input and generates current spikes, which serve as the cascading heat spike input of the neuristor B in subsequent layer. In this fashion, neuristor B integrates multiple heat spikes from neuristor A and fires a spike. Bottom left: Voltage traces for neuristors A, B, with a 50% duty cycle square input pulses. Neuristor A exhibits leaky integration of the input electrical pulses, while neuristor B performs similar integration of heat spikes from neuristor A, eliminating the need for complex buffer circuits. Bottom right: Similar configuration but with 80% duty cycle square input pulses.

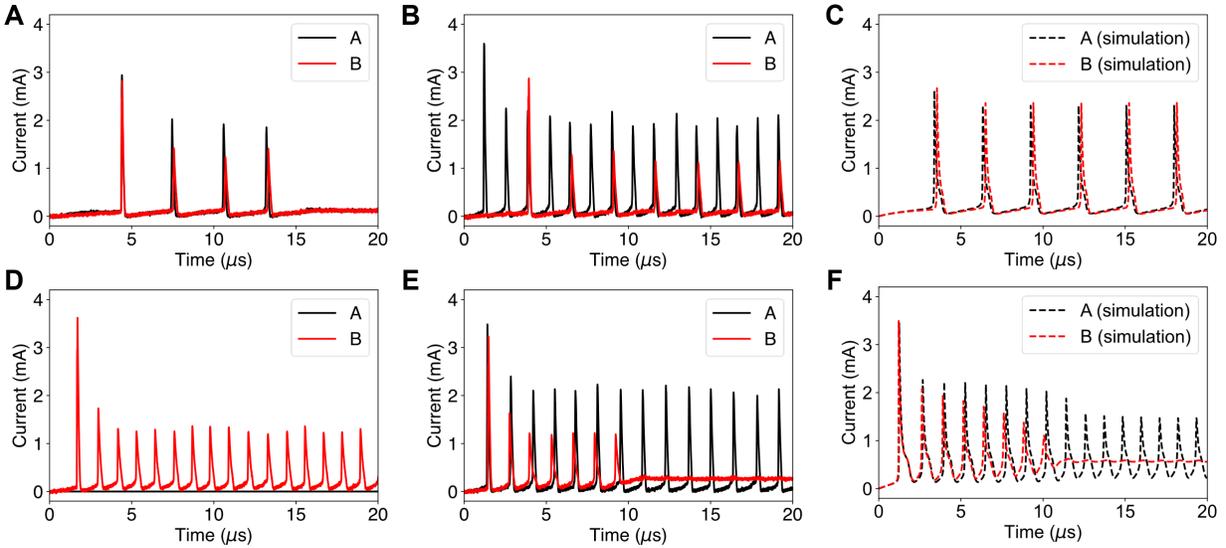

**Fig. 5. Excitatory and inhibitory interactions between neuristors.** Both neuristors are connected in series with a 12 $k\Omega$ load resistance. Neuristor A has an input threshold voltage of 2.9 V, while neuristor B has an input threshold voltage of 2.8 V. (A) When neuristor A is biased at 2.9 V and neuristor B is biased at 2.6 V subthreshold voltage, neuristor A excites neuristor B in a 1:1 excitation mode, resulting in both phase and frequency synchronization. (B) Neuristor A, at 4.3 V suprathreshold voltage, excites neuristor B (at 2.6 V subthreshold voltage) in a mixed-integer excitation mode with 3:1 and 2:1 spiking patterns. (C) Numerical simulation replicating the 1:1 excitation mode with subthreshold input voltages. (D) Neuristor B generates stable spikes when biased at 4.1V, with neuristor A inactive. (E) With both neuristors biased at 4.1V, neuristor B becomes quiescent after initial spikes due to inhibition by neuristor A. The stable heat spikes from neuristor A traps the neuristor B in the metallic state. (F) Numerical simulations mirroring the inhibitory behavior that neuristor A inhibits the stable spiking neuristor B, which is biased just below the upper threshold.

# Supplementary Materials for

## Reconfigurable cascaded thermal neuristors for neuromorphic computing

Erbin Qiu[*], Yuan-Hang Zhang, Massimiliano Di Ventra and Ivan K. Schuller

Corresponding author: Erbin Qiu e3qiu@eng.ucsd.edu

**The PDF file includes:**

Materials and Methods
Figs. S1 to S13
References

## *Materials and Methods*

Synthesis of epitaxial $VO_2$ on $Al_2O_3$ substrate

The $VO_2$ films were deposited on the (012)-oriented $Al_2O_3$ substrate by reactive RF magnetron sputtering. Initially, the (012) oriented $Al_2O_3$ substrate was loaded into a high vacuum chamber with a base pressure of ~ $1 \times 10^{-7}$ Torr. The sample holder was heated to 680 °C. Then pure argon was flown into the chamber at 2.2 s.c.c.m and 2.1 s.c.c.m mixed gases (20% oxygen and 80% argon). The sputtering plasma was triggered at a pressure of 4.2 mTorr by applying a forward power of 100 W to the target, which corresponded to an applied voltage of approximately 240 V. The deposition of $VO_2$ films lasted 30 mins, achieving a thickness of 100 nm. Upon completion, the sample holder was cooled down to room temperature at a 12 °C/min rate.

$VO_2$ crystal structure

The crystal structure of the $VO_2$ thin film was confirmed with a Rigaku Smartlab using 2-theta/omega scan from 20 deg. to 100 deg. in 0.01 deg steps. The Smartlab has a high resolution of 0.0001° and the characteristic wavelength of the copper X-ray tube is 1.5406 Angstroms.

Fabrication of $VO_2$ neuristor arrays

A Vistec (100kV) Electron Beam Lithography system was used to pattern $VO_2$ neuristor arrays. Each neuristor has dimensions of $100 \times 500$ nm$^2$ and is separated by 500 nm gaps. The e-beam resist, PMMA-A4, was spin-coated onto samples measuring 10 mm x 10 mm followed by a baking process at 115 °C for 20 mins for the first lithography step. Electrodes were defined by depositing a 15 nm Ti layer followed by a 40 nm Au layer. To investigate the thermal interaction between two neuristors, a second lithography and etching process were required. The negative e-beam resist ma-N 2405 was spin-coated onto the previously prepared samples and subsequently baked for 1 min at 91 °C for the second lithography step. An Oxford Plasmalab 80 Plus RIE system was used to etch the uncovered $VO_2$ films between the devices, while the negative resist protected the covered electrodes and the devices from being etched away.

Fast electrical dynamics measurements

A Tektronix Dual Channel Arbitrary Function Generator, AFG 3252C, was used to apply DC or pulse voltage bursts to the circuit. The AFG 3252C offers precise waveforms with both fast leading and trailing times of 2.5ns. The dual-channel feature allows us to control the device under test (DUT) individually while ensuring synchronized output signals to the circuit.

For recording ultrafast electrical dynamic signals, we employed the Tektronix Oscilloscope MSO54, which offers a maximum bandwidth of 1GHz and sampling rates of 6.25 GS/s. In particular, the channel impedance for measuring the voltage dynamics was configured to 1 MΩ, while the channel impedance for measuring the spiking current dynamics was set to 50 Ω.

Details of numerical simulations

*Model of the VO₂ hysteresis loop.* We employed the hysteresis model developed in (44), which assumes that in the insulating phase, VO₂ functions as an inherent semiconductor, characterized by an exponential R-T curve. Conversely, the metallic phase of VO₂ exhibits constant resistance. The metallic phase fraction of VO₂ undergoes a smooth transition from 0 to 1 across a critical temperature, which is different for the heating and cooling branches. Without delving into the derivations, we present the resulting equations:

$$R(T) = R_0 \exp\left(\frac{E_a}{T}\right) F(T) + R_m \quad (S1)$$

$$F(T) = \frac{1}{2} + \frac{1}{2}\tanh\left(\beta\left\{\delta\frac{w}{2} + T_c - \left[T + T_{pr}P\left(\frac{T - T_r}{T_{pr}}\right)\right]\right\}\right) \quad (S2)$$

$$T_{pr} = \delta\frac{w}{2} + T_c - \frac{1}{\beta}[2F(T_r) - 1] - T_r \quad (S3)$$

$$P(x) = \frac{1}{2}(1 - \sin\gamma x)[1 + \tanh(\pi^2 - 2\pi x)] \quad (S4)$$

In Eq. (S1), $R_0 \exp\left(\frac{E_a}{T}\right)$ represents the resistance of the insulating phase, $R_m$ denotes the resistance of the metallic phase, and $F(T)$ is the volume fraction of the insulating phase. $F(T)$ is defined by Eq. (S2), exhibiting a smooth transition from 1 to 0. In Eq. (S2), $\beta$ serves as a fitting parameter, $\delta$ takes the value of 1 in the heating branch and −1 in the cooling branch, $w$ is the width of the hysteresis loop, and $T_c$ is the critical temperature. $T_{pr}$ is the proximity temperature at the reversal point, which is introduced to characterize the distance from the current point to the major loop and is essential for characterizing minor and nested loops. $P$ stands for the proximity function, which is an arbitrarily chosen, monotonically decreasing function, introduced to approximate the proximity temperature at an arbitrary point. $T_{pr}$ is further defined in Eq. (S3), in which $T_r$ is the reversal temperature. Additionally, $P(x)$ is provided in Eq. (S4), with $\gamma$ as another fitting parameter. A detailed description and explanation of Eqs. (S1) - (S4) is described in reference (44).

We conducted experimental measurements of the R-T curves for our VO₂ sample using various heating and cooling cycles. The results are illustrated in Fig. 1C in the main text. To estimate the optimal parameters for the model, we employed the differential evolution algorithm (45), as implemented in the SciPy library (46), to fit the experimental data to Eqs. (S1) - (S4). The results are: $R_0 = 5.359 \times 10^{-3}\Omega$, $R_m = 262.5\Omega$, $E_a = 5220K$, $\beta = 0.253K^{-1}$, $w = 7.193K$, $T_c = 332.8K$, and $\gamma = 0.956$.

Note that the measured resistance curves contain non-ideal factors such as contact resistance between VO₂ and the electrodes. For comparison, the conductivity of metallic VO₂ is $8 \times 10^5 S/m$ (47), or 2.5 Ω for our sample of dimensions 100 nm × 500 nm × 100 nm. Given the fitted $R_m = 262.5\Omega$, this indicates that the contact resistance is the primary contributor in the metallic state. On the other hand, for the insulating state, the activation energy of VO₂ thin films at 20°C is about 0.22 eV (44), which gives a theoretical value of $E_a = 2553K$, about half of the fitted result. Indeed, we can observe a divergence between data and model in the insulating state as depicted in Fig. 1C, suggesting some limitation in our model. However, it is important to note that this model does accurately reproduce the R-T curves near the insulator-to-metal transition, which is our principal area of interest.

*Model of a single neuristor.* The circuit configuration is shown in Fig. 1B in the main text, where the VO$_2$ nanodevice is modeled as a temperature-dependent resistor in parallel with an intrinsic parasitic capacitance. A variable load resistor is connected in series with the VO$_2$ nanodevice, whose resistance is typically comparable to the insulating state of the VO$_2$ nanodevice.

The model of the hysteretic behavior of VO$_2$ is already illustrated in the previous section. However, it is worth noting that the hysteresis loop in Fig. 1C is measured through a quasi-static process, where the device undergoes slow heating, ensuring the entire sample transitions completely into the metallic state above the transition temperature. On the other hand, the scenario in a spiking oscillator is markedly different. As demonstrated in (39), narrow metallic channels emerge within the insulating bulk during each current spike, which yields a higher resistance relative to the fully metallic state recorded in the quasi-static process. In practice, to bypass the intricate task of modeling the complex nonequilibrium thermodynamics, we assume that the metallic channel has a volume fraction of $1/k$, where $k$ is a numerically fitted parameter. In other words, the resistance of the metallic state increases to $k$ times the value measured through the quasi-static process. Then, thermal conduction is modeled assuming uniform temperature within the neuristor and a constant environmental temperature.

Combining the circuit model and thermal model, we arrive at the following equations:

$$\frac{dV_1}{dt} = \frac{V_{in}}{R_{load}C} - V_1 \left( \frac{1}{R_{VO_2}C} + \frac{1}{R_{load}C} \right) \qquad (S5)$$

$$\frac{dT}{dt} = \frac{V_1^2}{R_{VO_2}C_{th}} - \frac{S_{th}(T - T_0)}{C_{th}} \qquad (S6)$$

Here, $V_1$ represents the voltage across the neuristor, $V_{in}$ is the input voltage, $R_{VO_2}$ refers to the resistance of VO$_2$, $C$ is the parasitic capacitance, $T$ denotes the temperature of VO$_2$, $T_0$ denotes the environment temperature, and $S_{th}$ and $C_{th}$ stands for the thermal conductance and thermal capacitance of the neuristor, respectively.

The resistance of VO$_2$ is slightly altered from the quasi-static hysteresis behavior in Eq. (S1) and is expressed as:

$$R_{VO2}(T) = R_0 \exp\left(\frac{E_a}{T}\right) F(T) + kR_m \qquad (S7)$$

where an additional factor, $k$, is introduced in Eq. (S7). This factor accounts for the formation of thin metallic channels during the spiking dynamics.

To precisely emulate the spiking behaviors, we extracted key features from the experimentally measured I-t curves and employed the numerical model to replicate these features. Specifically, the characteristics utilized in the numerical fittings include the amplitudes and positions of the first three spikes, the average width of the spikes, and the frequency of spiking. Once again, the differential evolution algorithm was employed to optimize the parameters, yielding the following results: $C = 145pF$, $k = 4.90$, $S_{th} = 0.206mW/K$ and $C_{th} = 49.6pJ/K$. These lead to an excellent agreement between the experimental and simulated curves as shown in Fig 2.

It is important to mention that we operated under the simplified assumption of a uniform temperature across the neuristor and a constant environment temperature. Therefore, the thermal conductance and capacitance calculated here are equivalent values considering the entirety of the device, which includes the VO$_2$, the electrodes, and the surrounding substrates. For comparison, the heat capacity per volume of VO$_2$ at 336K is $3.07 \times 10^6 \, J \, m^{-3} \, K^{-1}$ (48). For our sample with dimensions 100 nm × 500 nm × 100 nm, the heat capacitance is a mere 15.2 fJ/K, representing only 1/3000 of the fitted value. This demonstrates that the heat capacitance of VO$_2$ constitutes only a minuscule portion of the neuristor.

*Model of thermal coupling between adjacent neuristors.* To account for the thermal coupling between two adjacent neuristors, we introduce a coupling term to Eq. (S6):

$$\frac{dT_1}{dt} = \frac{V_1^2}{R_{VO_2,1}C_{th}} - \frac{S_{th}^{env}(T_1 - T_0)}{C_{th}} - \frac{S_{th}^{12}(T_1 - T_2)}{C_{th}} \quad (S8)$$

$$\frac{dT_2}{dt} = \frac{V_2^2}{R_{VO_2,2}C_{th}} - \frac{S_{th}^{env}(T_2 - T_0)}{C_{th}} - \frac{S_{th}^{12}(T_2 - T_1)}{C_{th}} \quad (S9)$$

Here, the subscripts 1 and 2 denote the two neuristors. $S_{th}^{env}$ represents the thermal conductance between a neuristor and the environment, and $S_{th}^{12}$ represents the thermal conductance between the two neuristors.

We define the coupling strength, $\eta$, such that $S_{th}^{12} = \eta S_{th}$ and $S_{th}^{env} = (1 - \eta)S_{th}$. As depicted in Fig. 5C and 5F of the main text, this thermal coupling model effectively replicates the experimentally observed excitatory and inhibitory behaviors.

The coupling strength $\eta$ is hard to measure experimentally so we investigated it using numerical simulations. Fig. S11 illustrates the coupling between two neuristors under varying values of η. As the coupling strength increases, the neuristors exhibit a phase-locked, 1:1 spiking pattern. Conversely, when the coupling strength is decreased, neuristor B integrates a higher number of heat spikes from neuristor A, resulting in 2:1 or 3:1 spiking patterns, and the synchronization between the neuristors also weakens.
To delve deeper into the phenomenon of thermal coupling, we carried out supplementary simulations using the theoretical model, and the outcomes are depicted in Fig. S12 and S13.

Interestingly, phase shifts and interference patterns emerged in these simulations, though these phenomena are not as pronounced in experimental environments. When the spiking frequency of one neuristor is an integer multiple of the other's, a stable synchronization is achieved, which enhances excitatory interaction and reduces the likelihood of inhibitory interactions. Conversely, at non-matching frequencies, phase discrepancies arise, dampening the excitatory interactions while amplifying the inhibitory ones.

*Figs. S1 to S13*

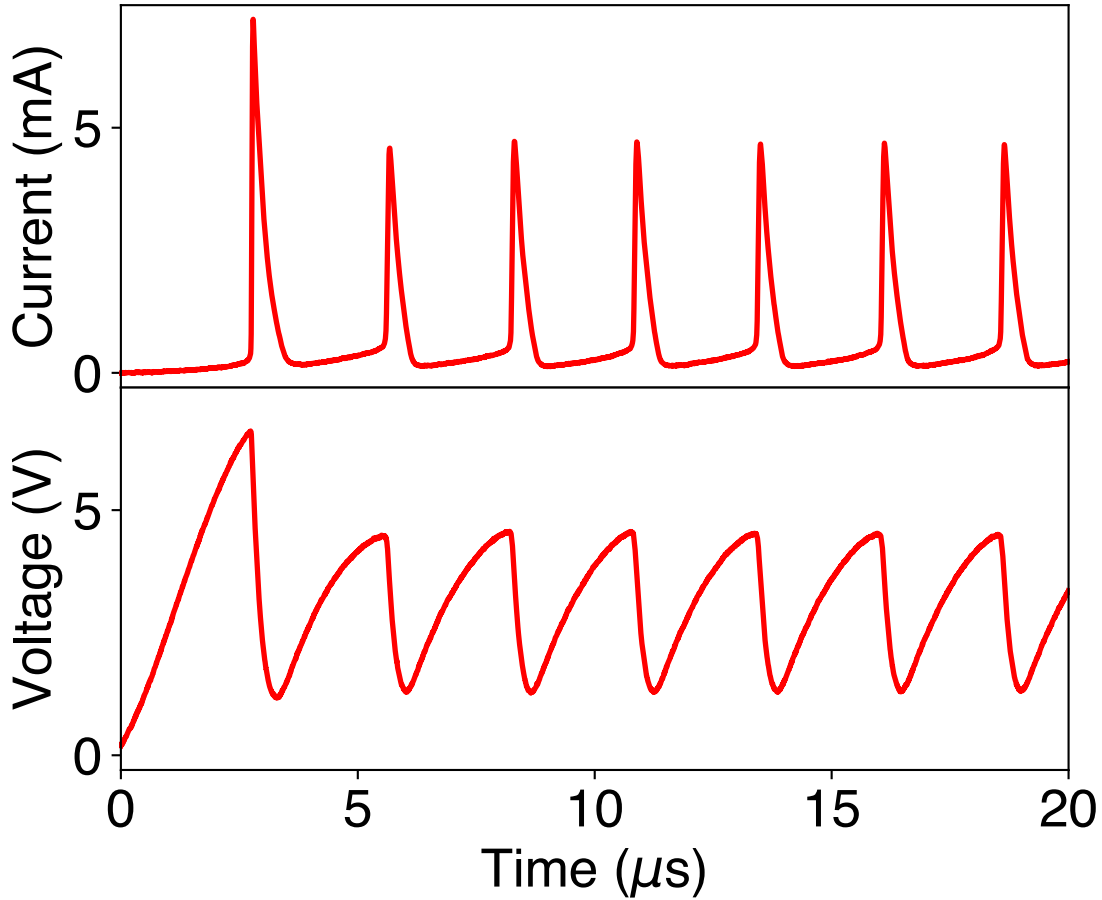

**Fig. S1. Typical Single Neuristor Behavior.** In this figure, the neuristor is connected in series with a 12 kΩ load resistor, and the input voltage is 12 V. When an input voltage is applied to the circuit, the ensuing current heats up the $VO_2$ nanodevice, concurrently charging up the parasitic capacitance. Upon reaching the critical threshold voltage, the VO2 undergoes an IMT. This abrupt decrease in $VO_2$'s resistance prompts the parasitic capacitance to discharge, resulting in a current spike. As the charge in the capacitance depletes, the majority of the input voltage is dropped across $R_{load}$, and the voltage across $VO_2$ does not generate sufficient heat to sustain the metallic state, causing the $VO_2$ to revert to its insulating state. This process repeats, generating a series of stable spiking auto-oscillations.

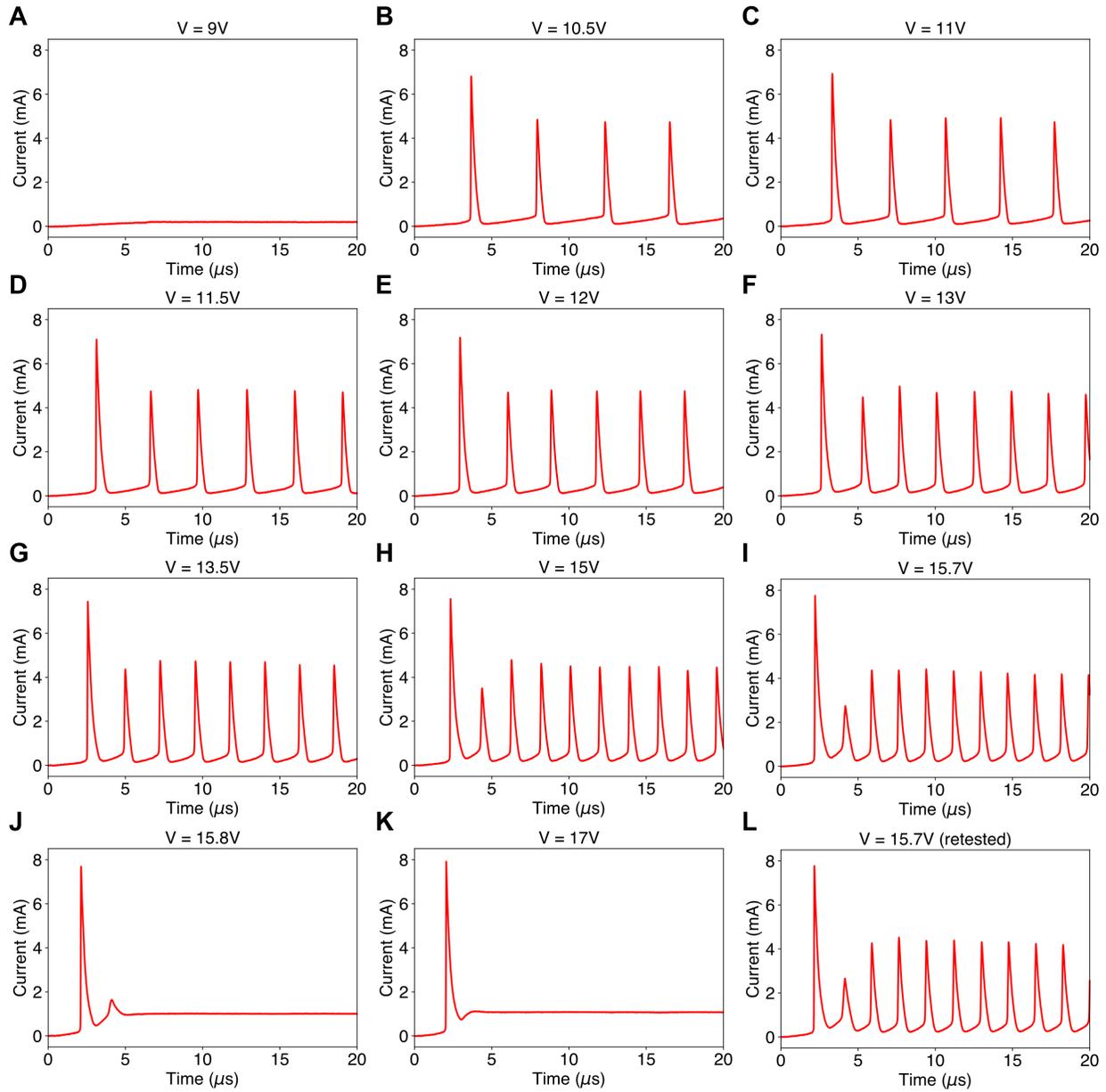

**Fig. S2. Comprehensive Analysis of Single Neuristor Behavior.** All the measurements presented in this figure were carried out at a baseline temperature of 325 K and with a load resistance set at 12 kΩ. The frequency response depicted in Fig. 2D of the main text is computed utilizing this data set collected using various voltage levels. (A) At a subthreshold voltage of 9V, the neuristor exhibits no current response. (B)-(I) When supplied with a sufficiently high input voltage, the neuristor operates as a spiking oscillator, with the spiking frequency increasing as the input voltage increases. (J)-(K) Upon the application of voltage exceeding the upper boundary, the neuristor becomes confined to its metallic state following a single current spike. (L) Subsequent to the experiments outlined in panels (J) and (K), the neuristor's behavior was rebiased at 15.7V, where stable spiking resumes. This confirms that the inhibitory behavior depicted in panels (J) and (K) is not due to degradation of the nanodevice.

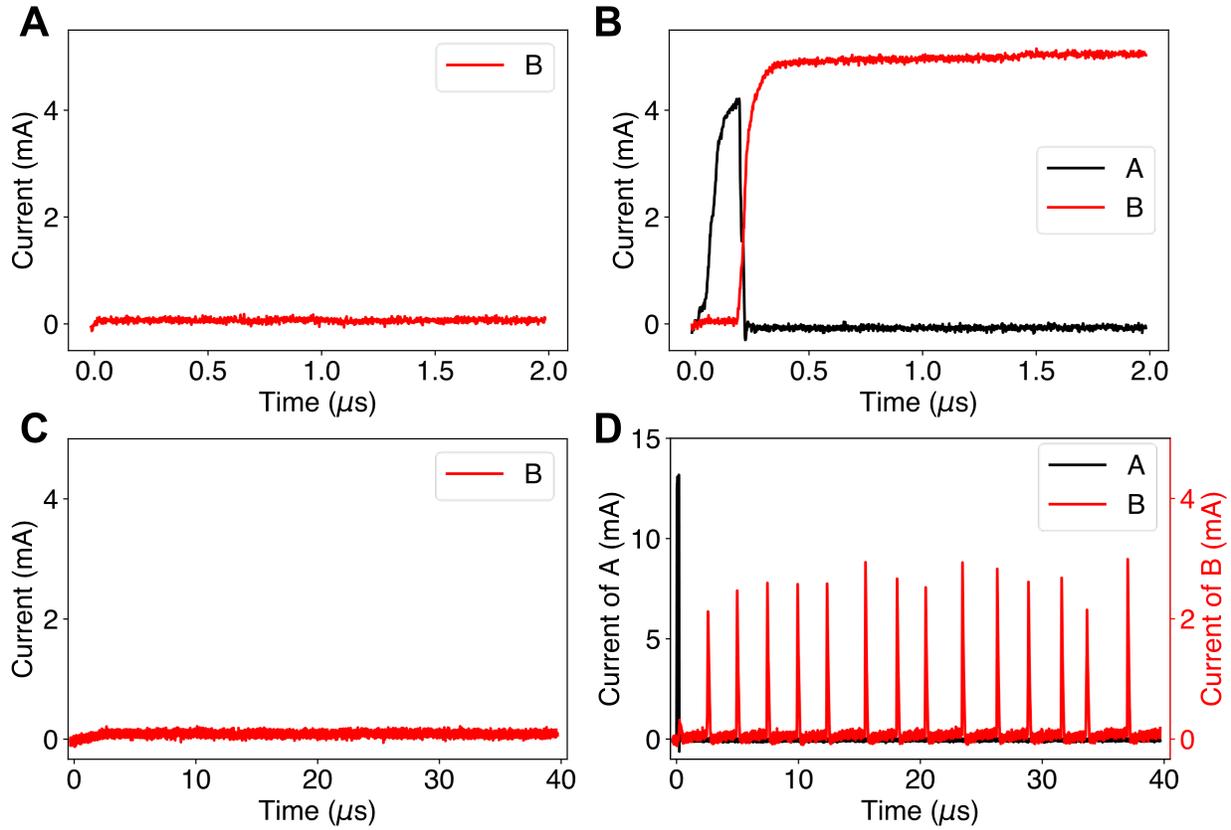

**Fig. S3. Baseline currents in coupled neuristors.** This figure serves as a supplement to Fig. 3A and 3B in the main text, showing the baseline currents. (A)(B) Spike-in and DC-out. Panel (A) demonstrates that neuristor B, biased with a subthreshold voltage of 1.5V, exhibits no current response when there is no input to neuristor A. However, as illustrated in panel (B), a current spike in neuristor A triggers the insulator-to-metal transition (IMT) in neuristor B, resulting in a direct current (DC) output. (C)(D) Spike in and spike out. Similarly, panel (C) indicates that neuristor B remains unresponsive when neuristor A is deactivated. However, a single spike in neuristor A initiates stable spiking patterns in neuristor B, as depicted in panel (D).

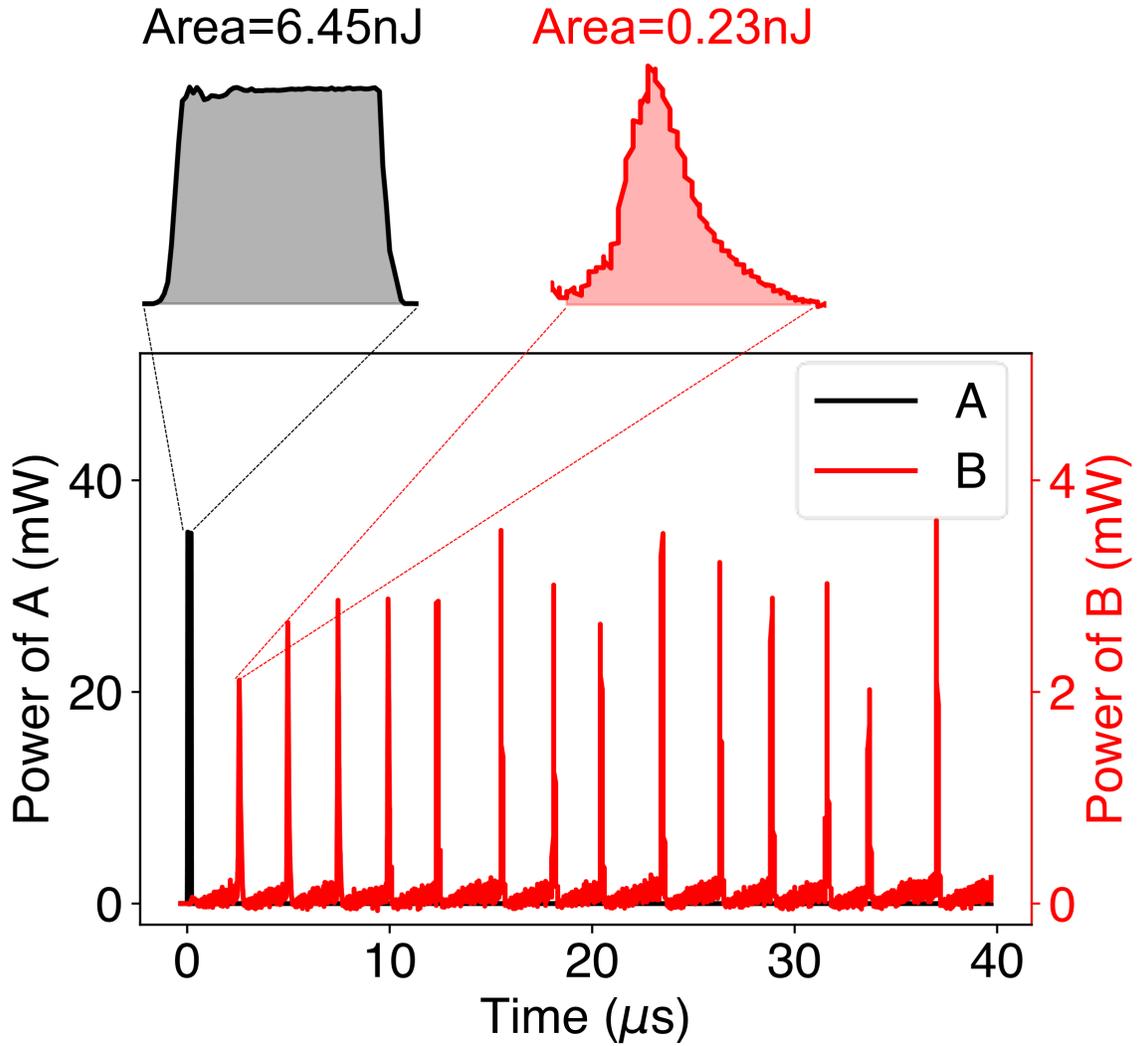

**Fig. S4. Power dissipation in coupled oscillators.** This figure is related to Fig. 3B in the main text and depicts the power dissipation over time in two coupled neuristors. The initial spike in neuristor A consumes 6.45 nJ of energy and activates 14 subsequent spikes in neuristor B. The energy consumption of the first spike in neuristor B is 0.23 nJ, and the total energy output for the 14 spikes is 5.56 nJ.

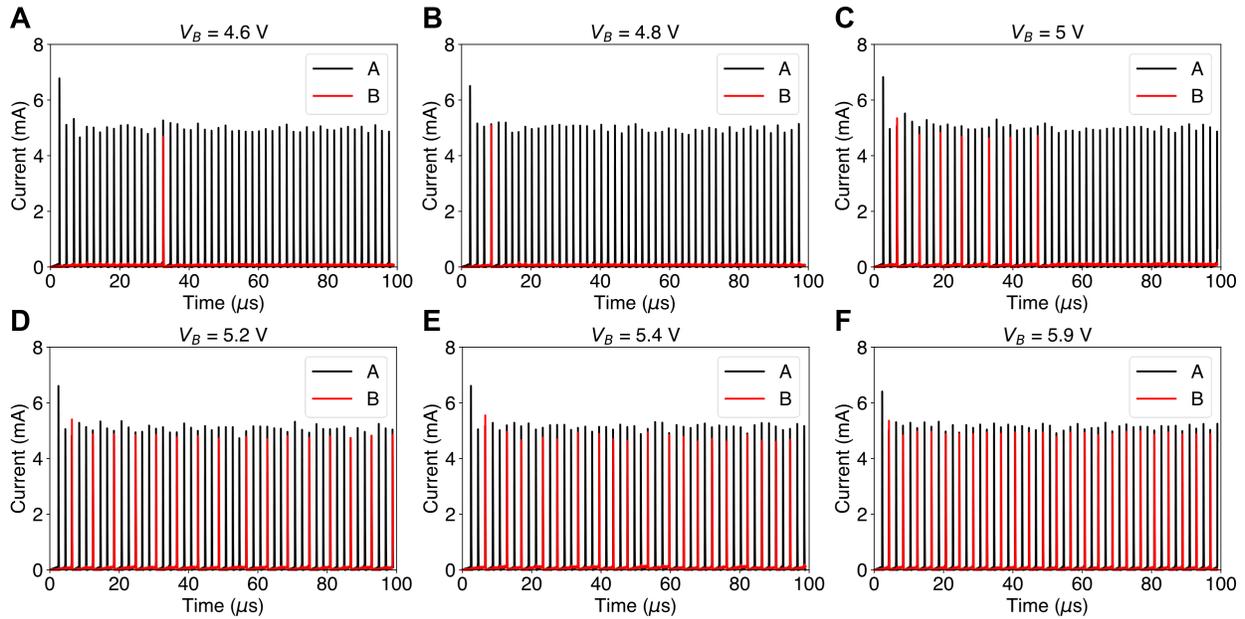

**Fig. S5. Stochastic leaky integrate-and-fire behavior in coupled neuristors.** This figure employs the same parameters as Fig. 3C in the main text, with neuristor A receiving an input of 7V in series with 22k ohms and neuristor B subjected to varying input voltages in series with 30k ohms. As the input voltage is increased, neuristor B needs to integrate fewer spikes from neuristor A, and the spiking pattern becomes more deterministic. This phenomenon has been systematically analyzed, and a corresponding heatmap is presented in Fig. 3D in the main text.

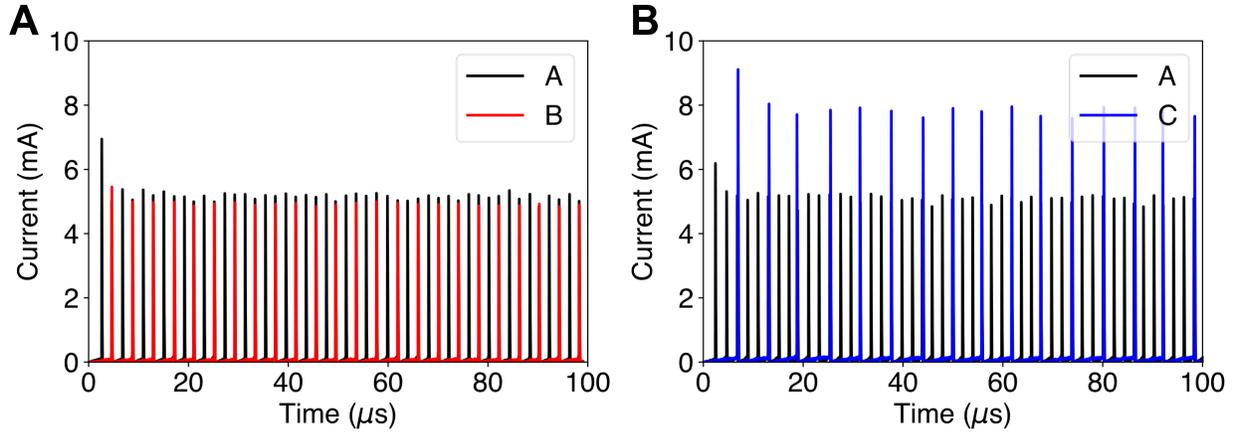

**Fig. S6. Impact of Distance on Information Transfer Among Neuristors.** In this figure, neuristors A, B, and C are positioned in parallel to each other, with a 0.5μm distance between A and B and a 1.5μm distance between A and C. All three neuristors have an identical threshold voltage of 6V. Neuristor A receives an input voltage of 7V, while neuristors B and C both receive input voltages of 5.9V. In this configuration, neuristor B needs to integrate two spikes from neuristor A to generate its own spike, whereas neuristor C needs to integrate three spikes.

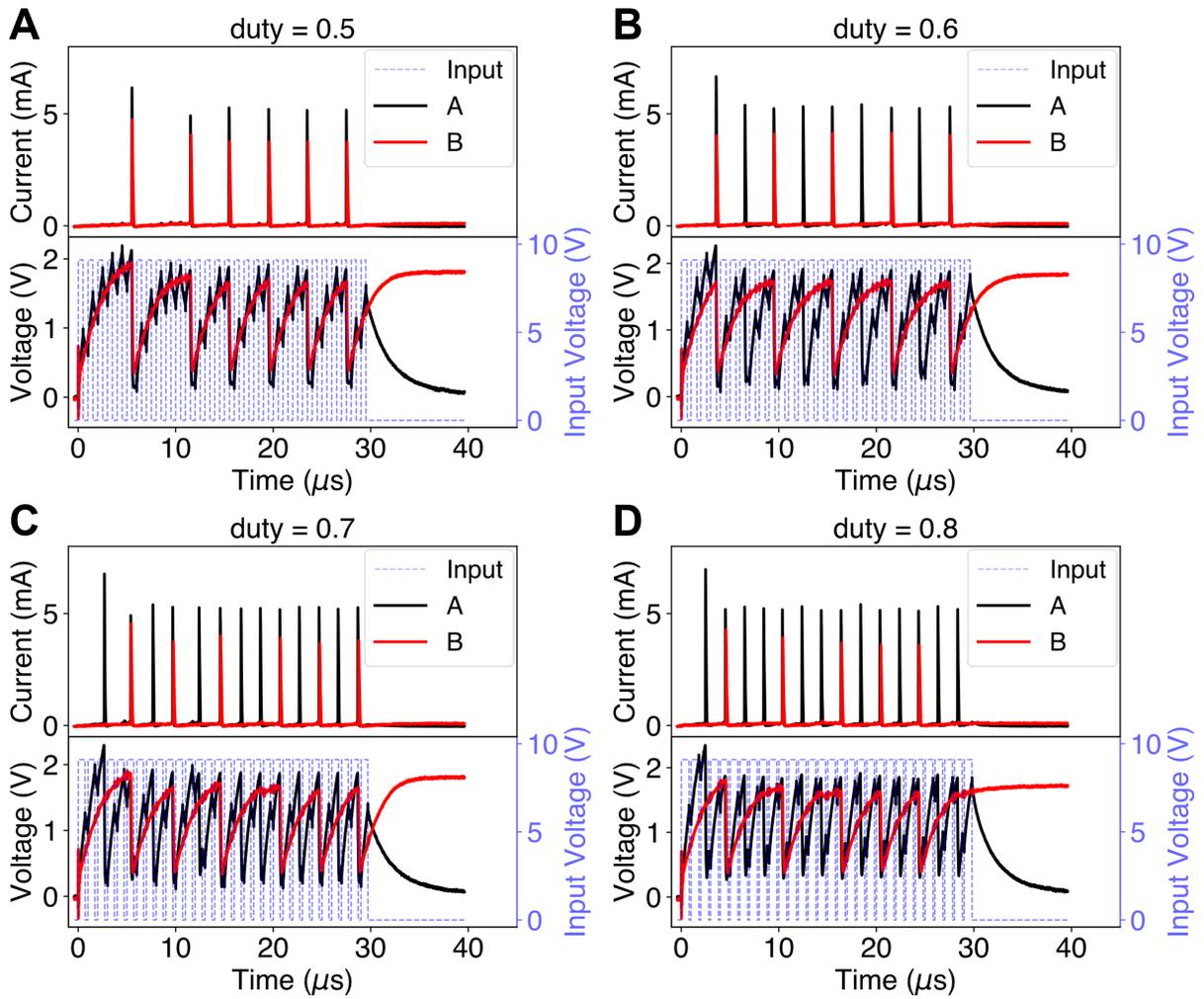

**Fig. S7. Cascaded information transfer between coupled neuristors.** This figure serves as an extension to Fig. 4 in the main text and illustrates the variations in spiking rates under different input duty cycles, as well as the relationship between current and voltage curves. In each panel, the sudden drop in voltage corresponds to a current spike. At a 50% duty cycle, neuristor A integrates more input electrical pulses, providing neuristor B with ample time for heat spike integration from neuristor A, and resulting in a 1:1 spiking pattern. As the duty cycle increases, the spiking frequency of neuristor A escalates. However, this leads to some spikes from neuristor A coinciding with neuristor B's refractory period, rendering them unable to trigger subsequent spikes in neuristor B. As a result, 2:1 or 3:1 spiking pattern emerges.

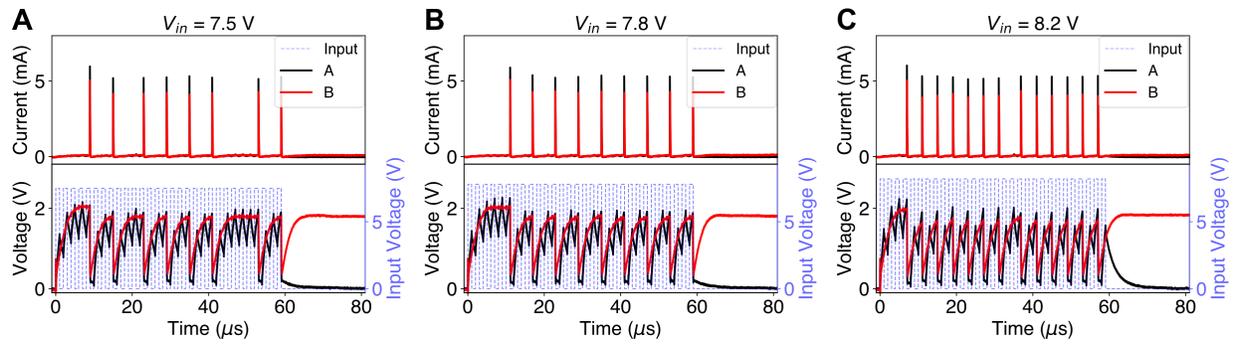

**Fig. S8. Cascaded Information Transfer in Coupled Neuristors with Varying Amplitudes.** This figure shares a similar setup with Fig. S6, with the distinction that neuristor A receives square pulses having a 50% duty cycle, a period of 2μs, and varying amplitudes. The rate coding behavior is evident, where an increased input voltage correlates with a higher frequency. This behavior is subsequently propagated through the following layers.

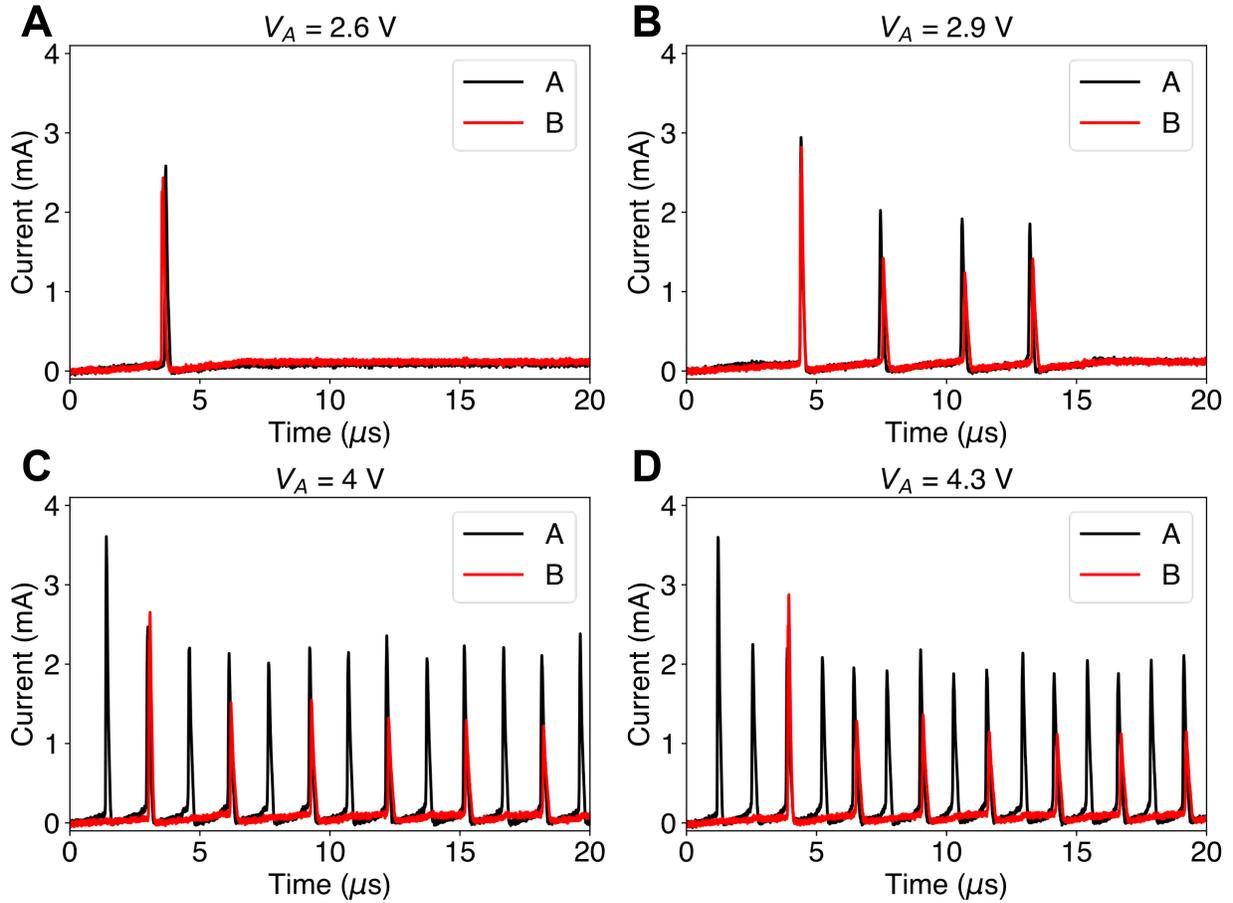

**Fig. S9. The evolution of excitatory characteristics with input voltage between adjacent neuristors.** This figure is an extension to Fig. 5 in the main text. Both neuristors are connected with a load resistance of 12kΩ. Neuristor A has a threshold voltage of 2.9V, whereas neuristor B has a threshold voltage of 2.8V. In all panels, the input voltage to neuristor B is kept constant at 2.6V, which is insufficient for it to spike independently. (A) With $V_A$ at 2.6V, the two neuristors mutually excite each other, generating a single spike. (B) At the threshold voltage $V_A = 2.9V$, every spike from neuristor A induces a corresponding spike in neuristor B, yielding a 1:1 spiking pattern. (C) At $V_A = 4V$, neuristor A exhibits a higher spiking frequency, leading to a 2:1 spiking pattern. (D) At $V_A = 4.3V$, the spiking frequency of neuristor A increases even further, while neuristor B adjusts to sustain the 2:1 spiking pattern.

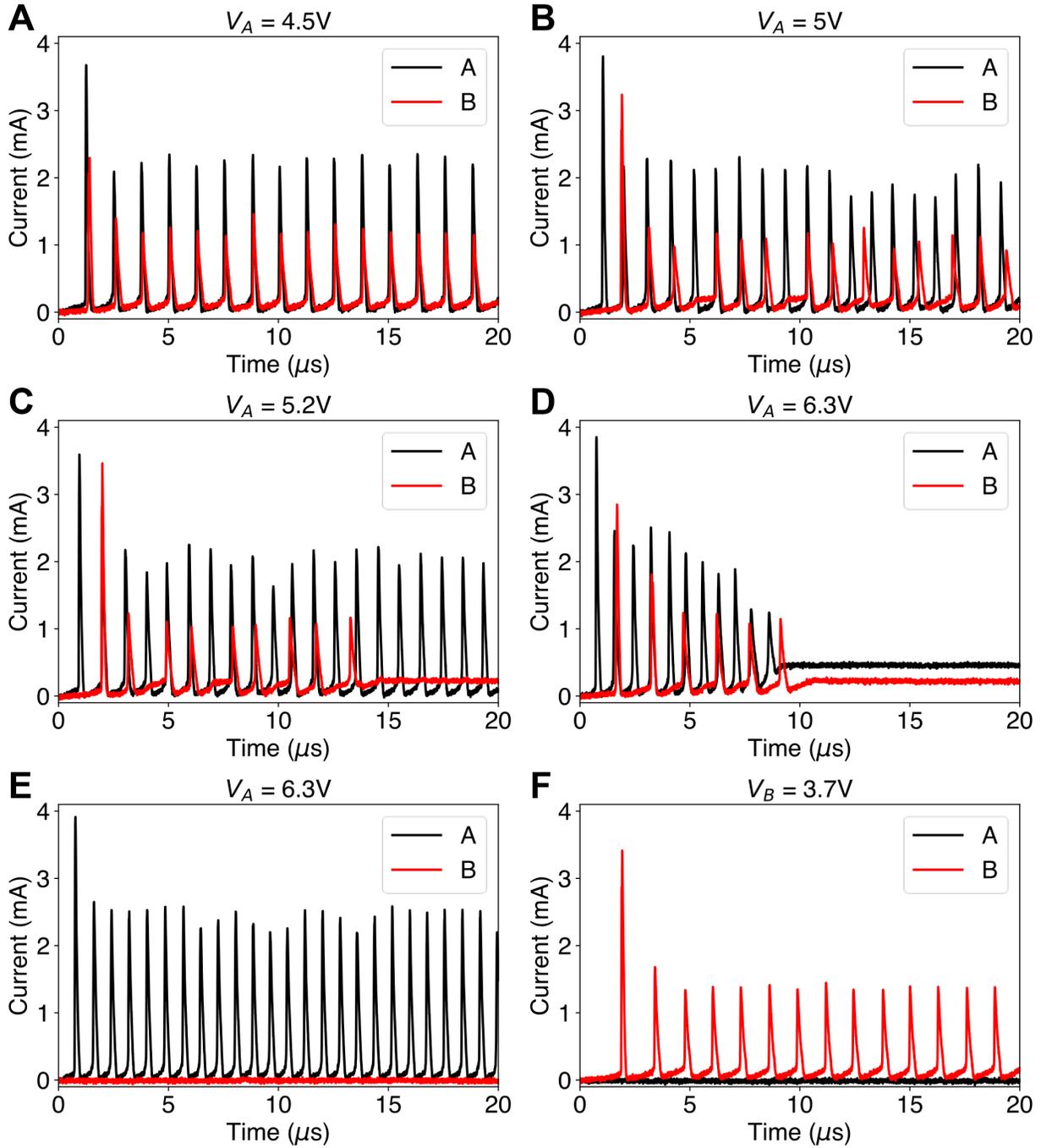

**Fig. S10. Another reconfigurable inhibitory characteristic between adjacent neuristors.** This figure also complements the results presented in Fig. 5 of the main text. Both neuristors are coupled with a load resistance of 12kΩ. Neuristor A possesses a threshold voltage of 2.9V, whereas neuristor B has a threshold voltage of 2.8V. Neuristor B's input voltage remains fixed at 3.7V, close to its upper boundary. (A) At $V_A = 4.5V$, the neuristors synchronize, displaying a 1:1 spiking pattern. (B) Elevating $V_A$ to 5V increases neuristor A's spiking frequency, thereby disrupting the synchronization and causing irregular spiking amplitudes and phase mismatches in neuristor B. (C) Further increasing $V_A$ to 5.2V exacerbates the irregularities in neuristor B's spiking pattern to the extent that it eventually stops spiking. (D) With $V_A$ at 6.3V, mutual inhibition occurs between the neuristors, causing both to terminate spiking after their initial spikes. (E) Without input to neuristor B, neuristor A can spike independently with an input voltage of 6.3V. (F) Similarly, without input to neuristor A, neuristor B is capable of spiking independently at an input voltage of 3.7V.

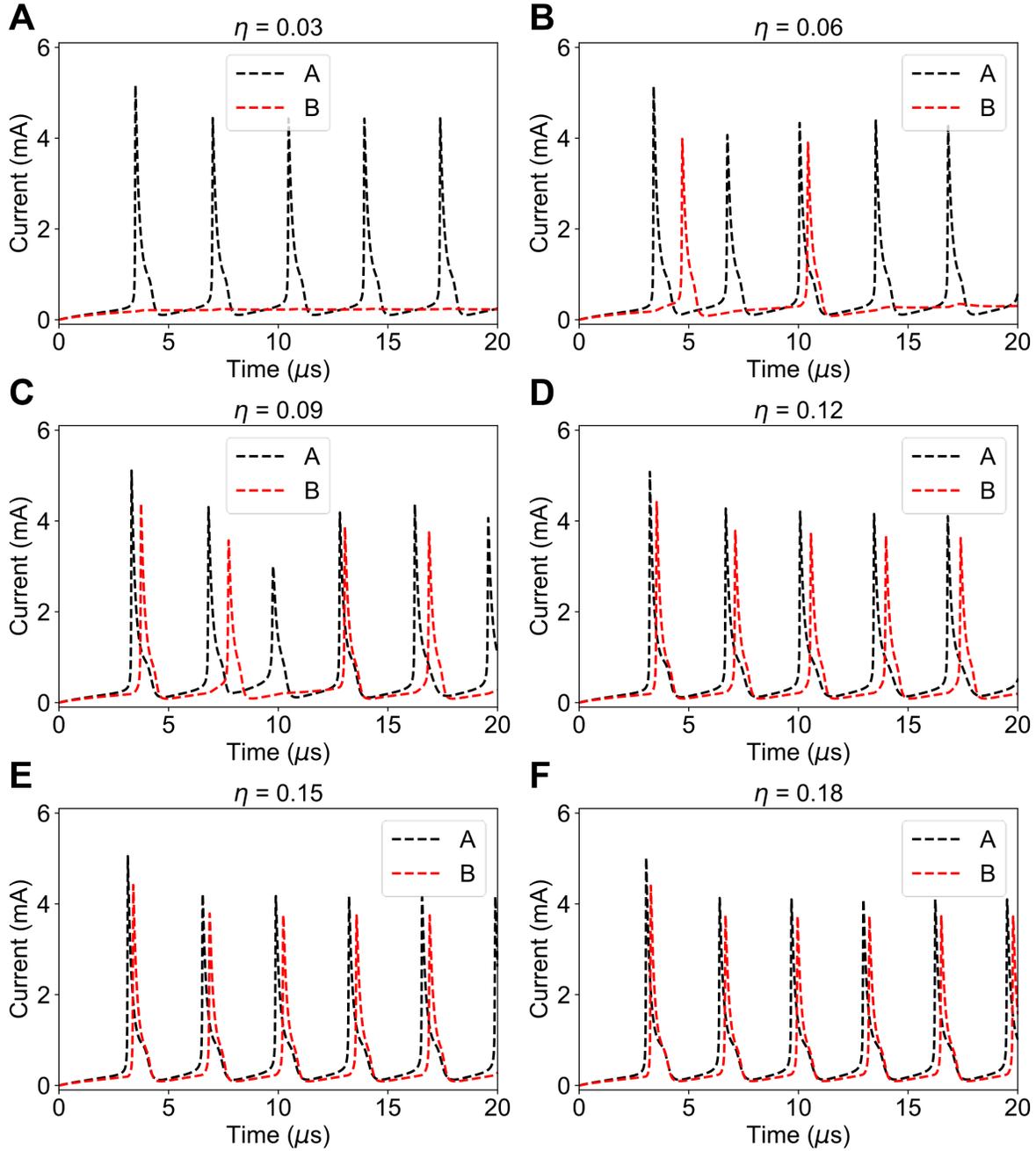

**Fig. S11. Numerical simulations illustrating excitatory behavior under various coupling strengths.** The parameters for the two neuristors in this simulation are the same as those outlined in Fig. 2 of the main text, with a spiking interval ranging from 10V to 15.7V when a 12kΩ load resistance is connected. The input voltages are set at $V_A = 11V$ and $V_B = 9.4V$, while the coupling strength $\eta$ between the two neuristors changes. (A) When $\eta = 0.03$, thermal energy from neuristor A is insufficient to induce spiking in neuristor B, which remains inactive. (B) At $\eta = 0.06$, spikes are excited within neuristor B, but the interaction is not strong enough to achieve synchronization. (C) With $\eta$ at 0.09, synchronization begins to emerge, albeit with occasional mismatches. (D)-(F) As $\eta$ continues to increase, the neuristors achieve stable 1:1 spiking synchronization, with the phase difference between them diminishing as η grows.

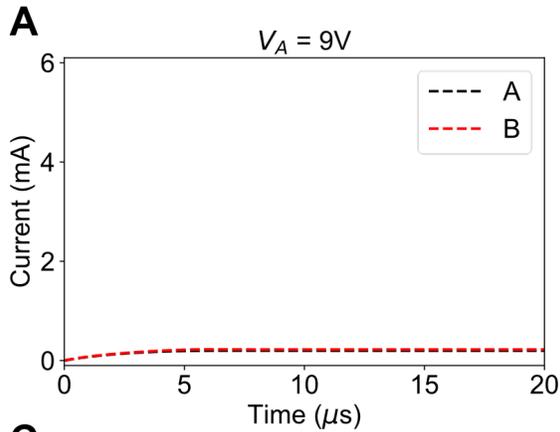
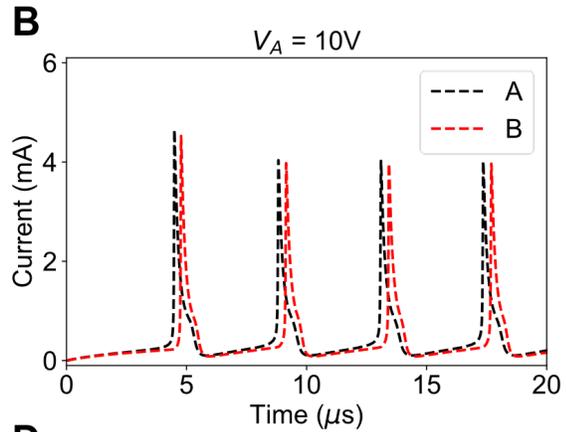
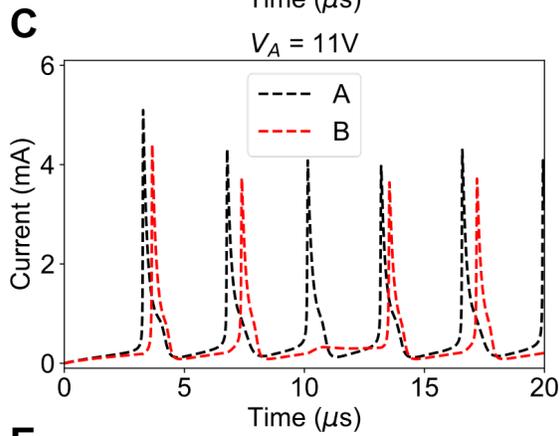
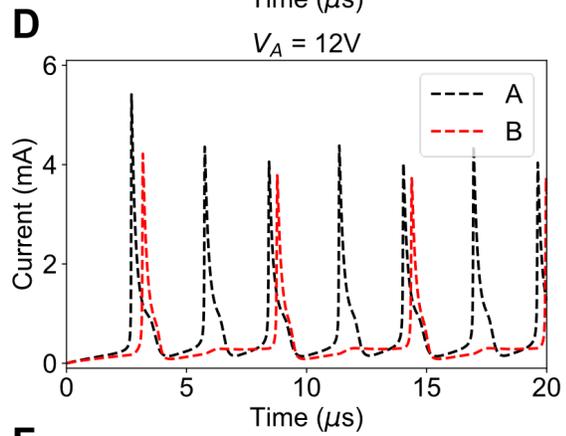
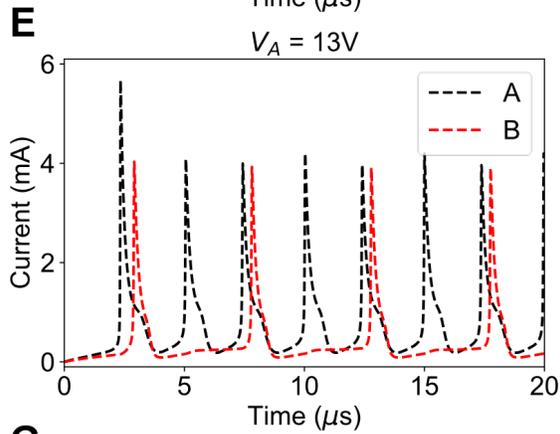
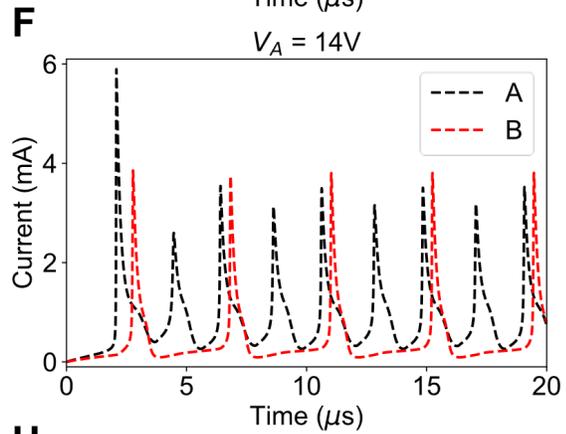
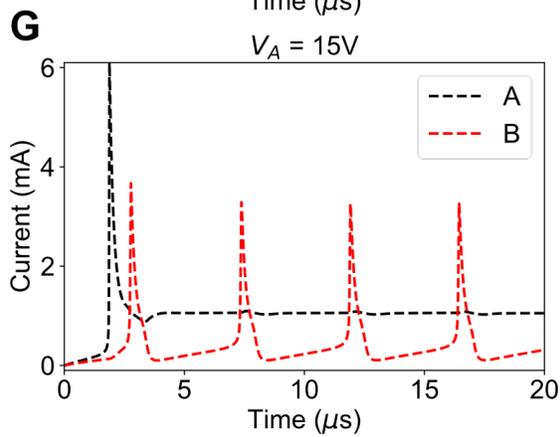
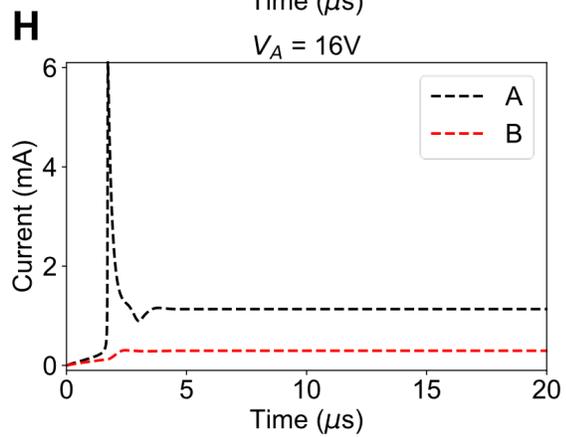

**Fig. S12. Numerical simulations of excitatory behavior under various input voltages.** This simulation uses a similar setting to Fig. S10, with a fixed coupling strength of $\eta = 0.1$ and $V_B = 9.4V$, while varying $V_A$. As depicted from panel (A) to panel (H), as $V_A$ increases, neuristor B transitions from a quiescent state to a 1:1 spiking mode, then to a 2:1 spiking mode, and ultimately reverts back to a quiescent state. When the frequency of neuristor A matches an integer multiple of the intrinsic frequency of neuristor B, pronounced synchronization is observed between the two neuristors. However, in cases where the frequencies do not align in this manner, phase mismatches occur, and the spiking amplitudes of the neuristors are slightly suppressed.

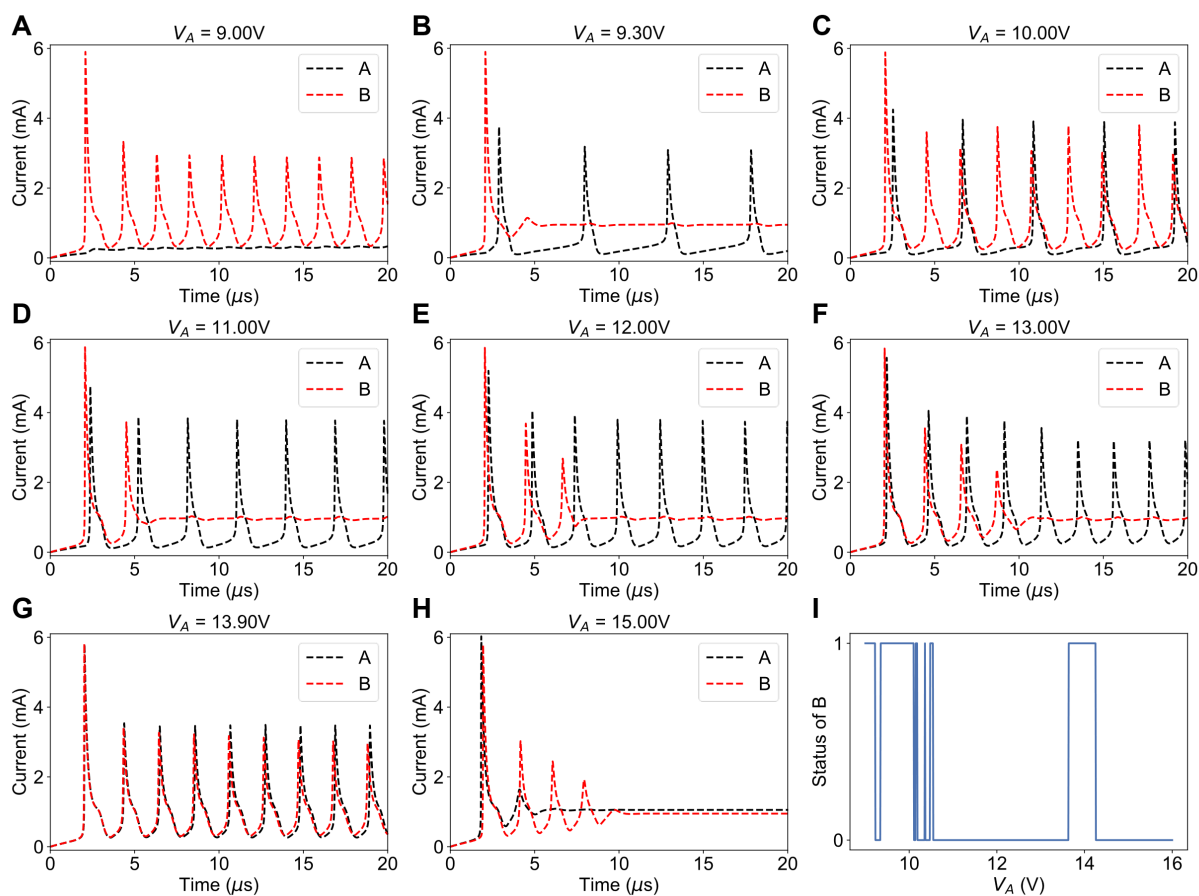

**Fig. S13. Numerical simulation of the inhibitory behavior under various input voltages.** This simulation also has a similar setting to Fig. S11, with a fixed coupling strength of $\eta = 0.1$ and $V_B = 14V$, while varying $V_A$. (A) With an input voltage of 9V, neuristor A does not spike, whereas neuristor B exhibits stable spiking at $V_B = 14V$. (B)-(H) A range of inhibitory interactions between the two neuristors. When neuristor A spikes stably, neuristor B is typically inhibited and ceases to spike after initial spikes. Notable exceptions are observed in panel (C), which shows a stable 2:1 spiking pattern, and panel (G), which features stable 1:1 spiking. This suggests that the inhibitory interaction stems from phase mismatches between the two neuristors. In panel (H), with $V_A$ set too high, mutual inhibition occurs between the neuristors. (I) This panel presents the state of neuristor B when different $V_A$ are applied. 1 representing spiking and 0 representing inhibition. The interaction proves to be rather complex; however, in general, neuristor B resumes spiking when the frequency of neuristor A is an integer multiple of its own.